\documentclass{sig-alternate-05-2015}

\usepackage{listings}
\usepackage[all]{nowidow}
\begin{document}






%

\title{T-Visor: A Hypervisor for Mixed Criticality Embedded Real-time System with Hardware Virtualization Support}
%
%
%
%
%

\numberofauthors{3} 
%
%
%
\author{\\
\alignauthor
Takumi Shimada\\
       \affaddr{The University of Tokyo}\\
       \affaddr{Tokyo, Japan}\\
       \email{shimada@is.s.u-tokyo.ac.jp}
\alignauthor
Takeshi Yashiro\\
       \affaddr{YRP Ubiquitous Networking Laboratory}\\
       \affaddr{Tokyo, Japan}\\
       \email{takeshi.yashiro@ubin.jp}
\alignauthor Ken Sakamura\\
       \affaddr{The University of Tokyo}\\
       \affaddr{Tokyo, Japan}\\
       \email{ken@sakamura-lab.org}
}

\maketitle
\begin{abstract}
Recently, embedded systems have not only requirements for hard real-time behavior and reliability,
but also diversified functional demands, such as network functions.
To satisfy these requirements, virtualization using hypervisors is promising for embedded systems.
However, as most of existing hypervisors are designed for general-purpose information processing systems,
they rely on large system stacks,
so that they are not suitable for mixed criticality embedded real-time systems.
Even in hypervisors designed for embedded systems,
their schedulers do not consider the diversity of real-time requirements and rapid change in scheduling theory.

We present the design and implementation of T-Visor, a hypervisor specialized for mixed criticality embedded real-time systems.
T-Visor supports ARM architecture
and realizes full virtualization using ARM Virtualization Extensions.
To guarantee real-time behavior,
T-Visor provides a flexible scheduling framework so that developers can select the most suitable scheduling algorithm for their systems.
Our evaluation showed that it performed better compared to Xen/ARM.
From these results, we conclude that our design and implementation are more suitable for embedded real-time systems than the existing hypervisors.
\end{abstract}

%
%

\ccsdesc[500]{Computer systems organization~Embedded software}
\ccsdesc[300]{Computer systems organization~Real-time system architecture}
\ccsdesc[100]{Computer systems organization~Real-time operating systems}

%
%

%
%
\printccsdesc


\keywords{virtualization; embedded system; real-time system}

\section{Introduction}
Hypervisors, also called virtual machine monitors (VMMs), are widely used in desktop and server environments today.
Basically, these are used for running multiple virtual machines (VMs) on one physical machine.
In the cloud systems, providers lend their VMs to multiple tenants.
This means different tenants may use the same physical servers.
VMs are independent, so that tenants are not affected by  each other even if they use the same physical server.

Hypervisors have also been applied to embedded systems recently.
However, their purposes are different.
One of their purposes is to construct mixed criticality systems.
Requirements in recent embedded systems are becoming more complex and various.
Traditionally, developers of embedded systems have employed real-time operating systems (RTOSes).
To satisfy recent complex requirements, general-purpose OSes (GPOSes), such as Linux, are also becoming popular in embedded systems.
However, complexity of GPOSes often leads to bugs and security vulnerabilities.
In addition, the majority of GPOSes do not consider real-time property.
Therefore, hypervisors are considered promising to allow co-existence of mixed criticality systems,
such as trusted RTOS and untrusted GPOS,
on the same physical platform.

Much work has been done so far to use hypervisors in embedded systems.
However, many problems have been left unsolved for practical use.
First, little consideration has been kept for minimizing vulnerabilities in hypervisors themselves.
In embedded systems, we can use some existing hypervisors designed for desktop and server environments,
such as KVM \cite{kvm:2014} and Xen \cite{xen_virt}.
Since they rely on GPOSes to provide rich functions, their trusted computing base (TCB) is large.
If there are vulnerabilities in hypervisors,
a high criticality VM can be affected by lower-criticality VMs.
To reduce such risks, we should keep TCB of hypervisors small in mixed criticality systems.

Second, there are often non-negligible cost to introduce virtualization layers.
Generally, applications in virtual environments tend to perform slower than those in bare-metal environments.
It is due to the overheads of virtual device drivers to share devices with other VMs,
interrupt latency caused by interceptions of hypervisors, and so on.

In addtiion, there are issues for real-time properties \cite{real-time:2012}.
How to schedule VMs is an important topic to assure hard real-time requirements.
General-purpose hypervisors usually use schedulers that do not consider real-time requirements.
Even if hypervisors are designed for embedded systems,
they often provide schedulers that have their own specific policies.
However, real-time scheduling techniques are often developed for each application or embedded system.
The best-fit scheduling algorithm largely depends from time to time on their requirements.
In addition, scheduling theory is becoming increasingly sophisticated.
For such diversity and rapid change,
the designs of schedulers of existing hypervisors are not considered optimal.

In this paper, we focus on mixed-criticality embedded real-time systems and implement a new hypervisor, called T-Visor.
We designed a hypervisor for ARM processors from scratch based on our prototype.
We have improved its implementation to run a GPOS as well as an RTOS.
We employ ARM Virtualization Extensions to realize full-virtualization with low overheads.
T-Visor provides independent virtual environments and unmodified OSes could run on them.
We designed a flexible scheduling framework of T-Visor to enable developers to implement their desirable algorithms.
Our implementation is much smaller than existing ARM hypervisors.
We evaluated T-Visor with an RTOS and a GPOS.
We compared our hypervisor with Xen using Linux benchmarks.
It revealed that the performance of T-Visor was better than that of Xen.

The main contributions of this paper are as follows:
\begin{itemize}
    \item To realize full-virtualization, we fixed bugs of virtual CPUs and implemented some features of virtual GIC,
        which previous version did not have.
    \item To realize full-virtualization, we implemented features which previous version did not have.
    \item We re-designed scheduling framework to become more flexible.
    \item We evaluated T-Visor using small benchmarks and real applications of Linux.
\end{itemize}


\section{ARM Architecture}
The ARM architecture is one of the famous embedded architectures.
To describe our implementation, we briefly introduce ARMv7-A architecture.
We also introduce the Virtualization Extensions, hardware support for virtualization available in ARMv7-A.

\subsection{CPU Mode}

\begin{figure}[!t]
\centering
\includegraphics[width=3.0in]{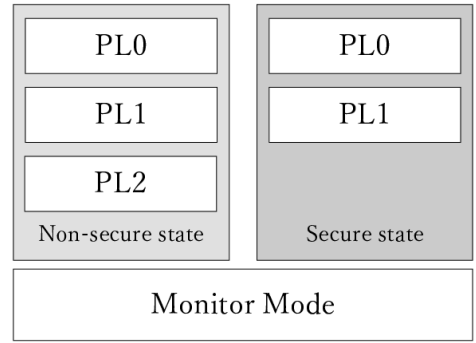}
\caption{ARMv7-A processor modes}
\label{fig_arm_mode}
\end{figure}

The overall view of  ARMv7-A processor modes is Figure \ref{fig_arm_mode}.
By the Security Extensions, which is previously called TrustZone and included in the Virtualization Extensions,
a processor has two security states, Non-secure state and Secure state.
These two states are isolated from each other.
In each states, all processor modes are available except Hyp mode.
Applications can switch security state through Monitor mode.

Processor modes are classified into privilege levels (PL).
PL0 is less privileged than PL1,
so that a processor in PL0 mode cannot access part of system registers.
Therefore, unprivileged applications on general OSes basically run on PL0
and sensitive functions in OSes run on PL1.
In ARMv7-A with the Virtualization Extensions, the special execution mode called Hyp mode was added.
Hyp mode has higher privilege level (PL2) than normal modes (PL1\&0).
Basically, hypervisors run in Hyp mode and guest OSes run in PL1\&0.
Hyp mode has its own system registers which cannot be accessed from lower privilege levels.


A processor enters Hyp mode when Hyp trap exception is generated.
To generate Hyp trap exception arbitrarily, Hyp call is available in Non-secure PL1 mode.
Hyp Configuration Register (HCR) provides an interface for defining which operations are trapped to Hyp mode.
Interrupts also can be trapped to Hyp mode by setting HCR.

A processor starts in PL1 mode of Secure state.
Usually, a boot loader changes the state into Non-secure state and boot OSes.
\subsection{Memory Management}
\begin{figure}[!t]
\centering
\includegraphics[width=3.0in]{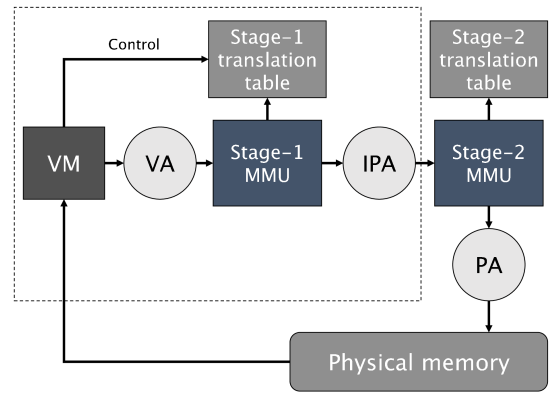}
\caption{Stage-1 MMU and Stage-2 MMU}
\label{fig_mmu}
\end{figure}
ARMv7-A has memory managed units (MMUs).
Figure \ref{fig_mmu} shows the overview of the address translation.
Usually, a MMU converts a virtual address (VA) into a physical address (PA).
Large Physical Adress Extension, which the Virtualization Extensions include, introduce Non-secure PL1\&0 Stage-2 MMU.
In Non-secure PL1\&0 mode, a program can use Non-secure PL1\&0 Stage-1 MMU as usual.
Stage-1 MMU translates a VA into an intermediate physical address (IPA).
An IPA is translated into a PA by Stage-2 MMU.
This translation does not need intervention of hypervisors.
Data abort related Stage-2 MMU can be routed to Hyp mode.

\subsection{Interrupt Management}
ARMv7-A processor has a built-in interrupt controller called Generic Interrupt Controller (GIC).
GIC is partitioned into the Distributor and CPU interfaces.

The Distributor prioritizes interrupts and distribute them to the specific CPU interface.
It also sends a software generated interrupt to CPU cores.
This can be used as an inter-processor interrupt (IPI).

The CPU interfaces determine whether interrupts have sufficient priority to cause exceptions.
ARMv7-A has two levels of interrupt exception request outputs, IRQ and FIQ.
The CPU interfaces also determine which level of exceptions are sent.
If an interrupt has sufficient priority, the CPU interfaces send the interrupt to the CPU cores.
The CPU can recognize the interrupt ID and CPU core ID which caused the interrupt via an interrupt acknowledge (ACK) register.
After handling interrupts, the CPU notifies the end of interrupt (EOI) through CPU interface.

GIC has virtualization support.
This introduces Virtual CPU interfaces.
The structure of Virtual CPU interfaces is the same as that of real CPU interfaces.
Therefore, hypervisors can control access to CPU interfaces by just redirecting them to Virtual CPU interfaces,
which helps hypervisors reduce Hyp trap of interrupt managing.
In Non-secure PL1\&0 mode, a program receives virtual interrupts instead of physical interrupts.
VMs also notify EOI through Virual CPU interfaces.
We can link this virtual EOI to physical EOI,
so that a program in Non-secure PL1\&0 can send physical EOI without Hyp traps.

However, the Distributor does not have virtualization support.
Therefore, hypervisors need to trap VM's accesses to the Distributor and emulate them.

\section{Design Principle}

The purpose of virtualization in embedded systems is different from that in cloud systems.
We focus on mixed criticality embedded real-time systems.

We define the requirements for our hypervisor as follows:
\begin{description}
    \item[Independency of VMs]
        In mixed criticality systems, trusted systems and untrusted systems co-exist on the same platform.
        Therefore, each VM must be independent, so that untrusted systems cannot affect trusted systems.
    \item[Small TCB]
        Large TCB often causes several security issues.
        We need to remove unnecessary large system resources to improve security.
    \item[Selectability of scheduling algorithms]
        There are several scheduling algorithms to satisfy real-time requirements in embedded systems.
        In general, the best scheduling algorithm depends from time to time on applications.
        Therefore, the scheduler should enable developers to select their algorithms.
    \item[Low performance degradation]
        Systems running on virtual machines are necessarily slower than those running on bare-metal environments.
        This performance degradation may affect real-time property.
        Therefore, we should carefully design the hypervisor to avoid this degradation.
    \item[Inter-VM communication mechanism]
        In cloud environments, inter-VM communication is not important because owners of VMs are different.
        However, in embedded systems, all VMs on the platform are generally owned by one tenant.
        Therefore, developers may need communication mechanism.
        For example, a GPOS providing user interfaces needs to tell commands to an RTOS.
\end{description}
Based on these requirements, we designed T-Visor.

\subsection{ARM Specific Design}
T-Visor considers only ARMv7-A architecture with Virtualization Extensions.
ARM architecture is one of major architectures for embedded systems.
In addition, this architecture has virtualization hardware to realize full-virtualization.
This helps T-Visor reduce various execution overheads and implementation complexity.
Therefore, we chose ARMv7-A architecture.

Supporting other major architectures, such as x86-based architectures, may be useful.
But these architectures often contain largely different points.
To avoid to increase implementation complexity, we did not consider other architectures.

\subsection{Type-1 Hypervisor}
\begin{figure}[!t]
\centering
\includegraphics[width=3.5in]{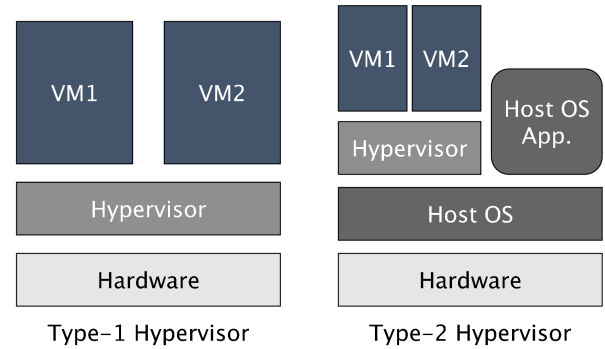}
\caption{Traditional Hypervisor Architecture}
\label{fig-hyp-arch}
\end{figure}
Traditionally, there are two types of hypervisor architecture: Type-1 and Type-2 (Figure \ref{fig-hyp-arch}).
We designed T-Visor as a Type-1 hypervisor.
A Type-1 hypervisor runs on the bare-metal environment.
Therefore, it need to implement startup code or device drivers.
In contrast, a Type-2 hypervisor can use its host OS functions.
This hypervisor does not need to have such hardware-specific functions because they are usually in the host OS.
However, TCB of a Type-2 hypervisor includes its host OS.
This does not satisfy one of our requirements, that TCB should be as small as possible.
In addition, the scheduling of the host OS may affect that of a Type-2 hypervisor.
For these reasons, we adopted the Type-1 design.

\subsection{Full-virtualization}
We adopted full-virtualization.
Full-virtualization has an advantage that developers can run unmodified OSes,
which makes it easier to run various types of OSes.
There are two ways to realize full-virtualization, one is to use hardware support and the other is dynamic binary translation (DBT).
We adopt the former way, as it is relatively low cost.
Today, there are few processors with virtualization support for embedded systems.
However, we believe that more embedded processors will have virtualization support in the near future.
The latter way does not need special hardware support.
Therefore, this way can be applied to wider range of processors.
However, its cost to implement is much higher than using hardware support.
High implementation cost tends to introduce software bugs.

Paravirtualization may be another alternative to realize our objective.
Existing approaches often adopt paravirtualization because they do not need hardware support.
However, this needs OS modification.
If developers use different types of OSes or different version of OSes,
they need to make different patches for each OS and verify that each OS runs correctly.

\subsection{Flexible Scheduling Framework}
There are a lot of real-time scheduling algorithms and techniques \cite{rt-survey:2012}.
Each algorithm or technique has its own advantages and disadvantages.
New approaches are also appearing year by year.
In addition, problems in scheduling VMs are different from scheduling processes in an OS.
For example, it is difficult to know internal states of VMs.
One of the solutions is modifying guest OSes to notify their states \cite{pas:2009}.
However, some developers may hesitate to use this because they do not want to modify guest OSes.
We designed the scheduling framework to enable developers to implement as various scheduling algorithms as possible.

\subsection{Pass-through Device Access}
We assumed that it is less important to share devices
among VMs in embedded systems than general purpose systems.
To share devices among VMs, we need to implement virtual device drivers.
Virtual device drivers involve not only implementation cost, but also execution overheads.
Therefore, we allowed most of device access pass-through the virtualization layer.
The hypervisor captures only access to forbidden devices.

\subsection{Static Resource Allocation}
T-Visor assigns physical resources to each VM at boot time statically.
This means each VM cannot use memory or devices which are not assigned at boot time.
The number of VMs is also fixed.
In usual embedded systems, roles of the systems are clearly defined at boot time.
Therefore, it can be assured that developers can predict how much resources each VM needs.
This helps to reduce implementation complexity of the hypervisor,
which makes the system TCB minimal.

\subsection{Keep the Systems Simple}
One of the important requirements of embedded systems is reliability.
Generally, it is difficult to verify systems are free from bugs.
One of the ways to reduce bugs is to reduce functions and modules.
Unnecessary functions or modules may involve errors and vulnerabilities.
The simplicity of the system helps developers check its correctness.
It may also enable to apply formal specification and verification.
Therefore, we did not employ any GPOSes or large libraries.

\section{Implementation}

Based on the design, we implemented T-Visor in C and ARM assembly.
T-Visor runs in Non-secure PL2 mode.
VMs run in Non-secure PL1\&0 mode.

T-Visor does not depend on any other systems except a boot loader
to load T-Visor and kernels of VMs into RAM.
Currently, we use U-Boot.
However, after booting, T-Visor does not use any function of U-Boot.
Therefore, we can use other boot loaders.

We now show the key feature of this implementation: providing virtual environments using ARM Virtualization Extensions (\S\ref{sec:virtual}),
GIC virtualization to manage interrupts (\S\ref{sec:interrupt}),
flexible scheduling framework (\S\ref{sec:sched}),
and inter-VM communication (\S\ref{sec:comm}).

\subsection{Virtual CPU}
\label{sec:virtual}
T-Visor provides virtual environments using ARM Virtualization Extensions.
Access to system control registers must be restricted not to affect the hypervisor or other VMs.
However, most of system control registers in PL0\&1 do not affect PL2.
Therefore, T-Visor does not implement traps for access to these registers except Auxiliary Control Register (ACTLR)
and cache and TCM lockdown registers (L2CTLR, L2ECTCLR).

Memory access is controlled by Stage-2 MMU.
If a VM tries to access forbidden memory area, it leads to Hyp trap.
Developers can configure handlers to these traps.
We implemented virtual GIC Distributor (the detail is explained in the next subsection) using this mechanism.

\subsection{Interrupt Control}
\label{sec:interrupt}
\begin{figure}[!t]
\centering
\includegraphics[width=3in]{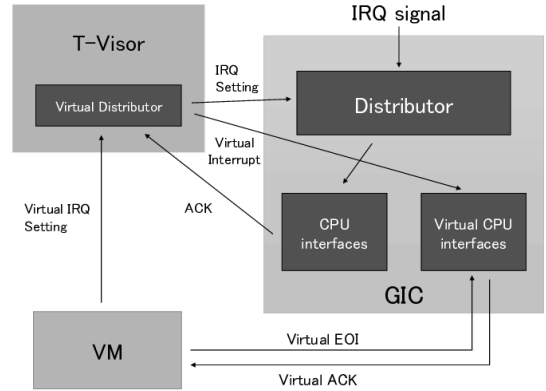}
\caption{Interrupt handling}
\label{fig:interrupt-control}
\end{figure}
When IRQ interrupt occurs, the processor enters PL2 mode.
T-Visor handles these interrupts.
VMs receive these interrupts as virtual interrupts.
VMs use virtual GIC CPU interfaces.
This reduces Hyp trap against access to CPU interfaces.
However, access to GIC Distributor must be trapped.
We implement a virtual GIC Distributor.
T-Visor captures access to GIC Distributor and emulate these access using the virtual GIC Distributor.
T-Visor uses this virtual distributor to send virtual interrupts.
Figure \ref{fig:interrupt-control} shows the overview of this mechanism.


\subsection{Flexible Scheduling Framework}
\label{sec:sched}

\begin{figure}[t]
\begin{lstlisting}[caption=The code of the function table,label=code-function-table]
typedef struct t_schedule_operation {
    void (*init)(void);
    T_VCPU *(*schedule)(void);
    void (*yield)(void);
    void (*block)(T_VCPU *);
    void (*unblock)(T_VCPU *);
    void *(*allocate)(T_VCPU *);
    void (*enque)(T_VCPU *);
} schedule_operation_t;
\end{lstlisting}
\end{figure}

To allow developers to implement various scheduling algorithms,
we provide scheduling function table inspired by that of RTEMS \cite{rtems:2014}.
From this table, the hypervisor calls functions on scheduling events.
Developers register their own functions to this function table.
The function table is defined as Code \ref{code-function-table}.

\texttt{T\_VCPU} is the structure for managing a VM.
This structure includes two parameters for scheduling,
one is for constant parameters (\texttt{sched\_param}),
the other is for dynamic states (\texttt{sched\_state}).
These parameters are held as void pointers.
After the initialization, the hypervisor uses these parameters only through the table functions.
Therefore, developers can pass pointers to arbitrary data structures and use these parameters arbitrarily in the scheduling functions.
For example, to implement a fixed-priority (FP) scheduler, a developer can use \texttt{sched\_param} as a fixed priority and \texttt{sched\_state} as a pointer to the wait queue element.

We briefly describe these functions of the function table.

\begin{description}
    \item[init] is an initialization function.
    This is called once in the boot time.
    \item[schedule] decides which VM to run next.
        This function itself does not manipulate states of T-Visor.
        A VM indicated by the return value should be executed next.
        If there is no executable VM, it returns a null pointer.

    \item[yield] function is called when the current executing VM enters sleep state.
        For example, a VM executes \texttt{wfi} instruction.
        If a VM yields, the VM must not be scheduled until the VM receives a wakeup event.

    \item[block] function is called when an executing VM are preempted by other VM.
        A blocked VM should be pushed to the scheduling wait queue.

    \item[unblock] function is called when a VM becomes executable.
        An unblocked VM should be pushed to the scheduling wait queue.

    \item[allocate] function provides a scheduling status parameter for \texttt{T\_VCPU}.
        \texttt{allocate} is called when \texttt{T\_VCPU} is constructed.
        \texttt{allocate} can use \texttt{sched\_param} in \texttt{T\_VCPU} to initialize the \texttt{sched\_state}.

    \item[enque] function is called to push a VM to the scheduling wait queue.
        Note, a structure of the wait queue can be an arbitrary data structure.
        A developer can use not only a priority queue, but also a simple list, a stack and so on.
        An enqueued VM is expected to be scheduled in the future.
\end{description}

Rescheduling will be done at the end of Hyp call and physical interrupt handling if rescheduling request flag is set.
This flag is set by user defined functions.
This helps developer control scheduling timing easily.

\subsection{Inter-VM Communication}
\label{sec:comm}
\begin{figure}[!t]
\centering
\includegraphics[width=3.5in]{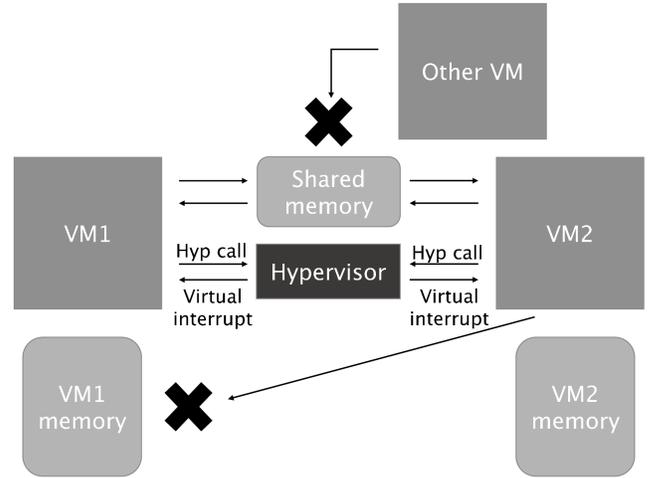}
\caption{The overview of communication between VMs}
\label{fig_comm}
\end{figure}
T-Visor provides an inter-VM communication mechanism built on shared memory pages and virtual interrupts.
Figure \ref{fig_comm} shows the overview of this mechanism.
In the initialization phase, T-Visor allocates some shared pages to VMs.
Each VM can access to the shared memory freely.
To avoid resource contention, each VM sends a virtual interrupt via T-Visor.

The shared memory is protected by the translation table.
Therefore, VMs which are not related to the communication cannot access the shared memory.
VMs cannot access other VM's memory outside the shared memory.
T-Visor sends a virtual interrupt which has an agreed-upon interrupt ID number.
Therefore, VMs receive only virtual interrupts that the developers permit.

VMs can remap arbitrary IPAs to the shared memory region.
The page size of shared memory, set in the translation table, is set 4KB.
Therefore, the shared memory is provided per 4KB and aligned to 4KB boundary.

This implementation permits access to shared memory without synchronization by virtual interrupt.
In our first implementation, we use a Hyp call to access the shared memory.
When a VM invokes a Hyp call to use a communication channel,
T-Visor checks whether the communication channel is available or not.
If the communication channel is available, Stage-2 translation table of the VM is modified to permit the access to the shared memory page.
This implementation prevents access to shared memory without synchronization.
However, modification of the translation table requires TLB flush.
The execution time of the benchmark program on the first implementation was about 10 times larger than that in the current implementation.

\section{Case Study of Implementing Schedulers}

In this section, we present a case study of implementing scheduler
to show flexibility of our scheduling framework.
We implemented Earliest Deadline First (EDF) scheduler on T-Visor.
We explain how to implement each function.

\subsection{Algorithm}
Each VM has period and execution time.
Every period, each VM can run until it uses up its execution time.

EDF scheduler gives the highest priority to a VM which has the eariest deadline.
In this algorithm, a deadline means time when a next period starts.

\subsection{Implementation}
We assigned a period and a execution time to each VM.
These parameters are given as \texttt{sched\_param} members.
Each VM has a dead line time and a remained execution time in \texttt{sched\_state}.
At first, a dead line of a VM is start time of the system plus its period.
This value is set in \texttt{allocate} function.

EDF scheduler has two priority queue.
One queue holds executable VMs.
In this priority queue, VMs are ordered by their dead line.
The other queue holds VMs waiting for next periods.
In this priority queue, VMs are ordered by their next periods.
These priority queue are initialized in \texttt{init} function.
In \texttt{enque}, a VM is pushed into the executable queue.

In \texttt{schedule} function, it checks waiting queue.
If the current time is over its next period starting,
a VM is pushed to the executable queue.
At that time, its \texttt{sched\_state} is updated to replenish the execution time.
Then, \texttt{schedule} function pop the top elements of the executable queue.
\texttt{schedule} function also set a timer event at the time a VM will expire its execution time.
This timer event calls \texttt{scheduler\_set\_flag} function, which lets T-Visor re-schedule.

In \texttt{block},
If a VM uses up execution time, it is pushed into the wait queue.
Otherwise, is is pushed into the executable queue.
This function also calls \texttt{scheduler\_set\_flag} function.

In \texttt{yield},
it just calls \texttt{scheduler\_set\_flag} function because a VM is not executable state.

In \texttt{unblock},
If a VM uses up execution time, it is pushed into the wait queue.
Otherwise, is is pushed into the executable queue.
This function also calls \texttt{scheduler\_set\_flag} function.

Through this implementation, we believe that our scheduling framework can support various algorithms.
Note, these functions we implemented do not modify status of VMs in T-Visor.
We completely separate these internal status.


\section{Evaluation}

To evaluate whether T-Visor achieves our requirements,
we conducted the experiments on Cubieboard2,
which has dual-core Cortex-A7 CPU, 1GB RAM, MHz Clock, and  100Mb Ethernet controller.
The compiler we used to compile T-Visor was GCC 5.3.0.
In these experiments, we used only one processor of Cubieboard2.

\subsection{Code Size}
\begin{table}[tb]
    \begin{center}
        \caption{LOC of T-Visor}
        \begin{tabular}{l|r}
            \hline \hline
            Language & LOC \\
            \hline \hline
            C Language & 5163 \\
            Assembly & 1067 \\
            \hline
            Total &  6230 \\
            \hline
        \end{tabular}
        \label{code_size}
    \end{center}
\end{table}
We counted LOC of T-Visor using SLOCCount \cite{sloccount}.
The result is shown in Table \ref{code_size}.
We also counted Xen 4.4.0 using SLOCCount.
Xen has support for other architectures.
Therefore, we only counted LOC of \texttt{xen/common} directory,
which does not include code depending on architectures.
The LOC of \texttt{xen/common} directory is 40,633.
Our code is about 6200 LOC in total.
This result showed that our design made the size of our implementation much smaller than that of Xen.

\subsection{Micro Benchmark}
\begin{table}[bt]
    \begin{center}
        \caption{Micro architectural latency}
        \begin{tabular}{l|r|r}
            \hline \hline
            & \multicolumn{1}{|c}{Time($\mu$sec)} & \multicolumn{1}{|c}{Cycle}\\
            \hline \hline
            Hyp call & 6.58 &  6000 \\
            Switch & 25.84 &  23564 \\
            Interrupt & 7.48 & 6824  \\
            Virtual interrupt & 29.71 & 27094 \\
            \hline \hline
        \end{tabular}
        \label{eval-micro}
    \end{center}
\end{table}

We measured latency of micro architectural operations to reveal the main reasons of overheads caused by T-Visor.
The result is shown in Table \ref{eval-micro}.

\texttt{Hyp call} is the time of executing an empty Hyp call.
This included the cost of register saving to enter T-Visor from VM and
register restoring to enter VM from T-Visor.

\texttt{Switch} is the cost of switching executing VM.
To measure this, we prepared a Hyp call which performs switching itself.
This included scheduling cost.
In this experiment, only two VMs run and T-Visor used a round-robin scheduler.

\texttt{Interrupt} is the overhead of handling an interrupt.
The time we measured was from starting the interrupt handler of T-Visor to the end of the interrupt handler of T-Visor.
The interrupt handler of T-Visor makes a virtual interrupt by writing LR register.
After the interrupt handler of the T-Visor, a VM restarts and processes an interrupt.

\texttt{Virtual Interrupt} is the overhead of emitting a virtual interrupt using Hyp call.
This included the cost of interrupt handling in a VM.
A virtual interrupt is employed in inter-VM communication.
Therefore, this is a rough estimation of the cost of inter-VM communication.

We believe that these latencies are small enough to run VMs.

\subsection{Linux benchmark}

\begin{figure}[!t]
\centering
\includegraphics[width=3.0in]{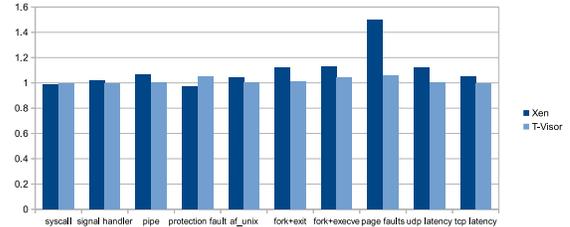}
\caption{lmbench result}
\label{fig-lmbench}
\end{figure}

We compared the performance of Linux on T-Visor with  DOM0 Linux of Xen 4.4.
We used Arch Linux with Linux 3.15.0 based kernel as a platform.
First, we run \texttt{lmbench3} benchmark suite \cite{lmbench:1996} to evaluate the overheads in a VM.
Second, we run some real application workloads.
Table \ref{app_desp} shows the detail of the workloads.
Basically, these applications are a subset of the workloads which was used in the evaluation of KVM/ARM \cite{kvm:2014}.
In these evaluations, we restricted the RAM size to 256MB and prepared 256MB swap area.
We showed these relative performance, the value normalized by bare-metal environment.

Figure \ref{fig-lmbench} shows the relative performance of lmbench programs.
In almost all benchmarks, T-Visor performed faster than Xen.
This shows our simple design reduced overheads in many cases.
In some benchmarks, Xen performed better than bare-metal and T-Visor.
The execution time of these benchmarks was much shorter than other benchmarks.
Therefore, it seemed in error ranges.

\begin{table}[!t]
    \begin{center}
        \caption{Real application workloads in Linux}
        \begin{tabular}{l|p{160pt}}
            \hline \hline
            Application & Description \\
            \hline \hline
            mysql & Evaluate MySQL v15.1 (10.1.12-MariaDB) using the SysBench \cite{sysbench} OLTP benchmark with the default configuration \\
            \hline
            memcached & Evaluate memcached v1.4.25 using the \texttt{memslap} benchmark with a concurrency parameter of 4 \\
            \hline
            curl 1G &  Run \texttt{curl} to download 1GB random data from the linux server running nginx in the same local network \\
            \hline
            untar & Running \texttt{tar} to extract Linux v4.4 sources compressed with bz2 compression \\
            \hline
            hackbench & Running \texttt{hackbench} \cite{hackbench} with 10 process groups running with 1000 loops to test scheduler and unix-sockets \\
            \hline \hline
        \end{tabular}
        \label{app_desp}
    \end{center}
\end{table}

\begin{figure}[!t]
\centering
\includegraphics[width=3.0in]{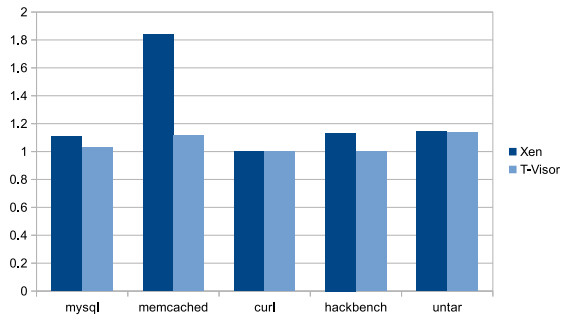}
\caption{Real application performance summary}
\label{fig-application}
\end{figure}

Figure \ref{fig-application} shows the relative performance of real application workloads.
In all workloads, T-Visor performed better than Xen.
Especially, \texttt{memcached} of Xen was much worse than that of T-Visor.
This application used much memory.
Therefore, it seemed to cause a lot of exception and use swap frequently.

\section{Related Works}
\subsection{Other Hypervisors}
Xen \cite{xen:2003} is one of famous Type-1 hypervisors.
Xen needs a special guest Linux, called DOM0.
DOM0 has virtual device drivers for VMs.
Before Virtualization Extensions are available, Xen can support ARM architectures using paravirtualization \cite{xen_arm:2008}.
Now, full virtualization is available in Xen on ARM architectures \cite{xen_virt}.

KVM is a Type-2 hypervisor.
KVM is available on ARM architecture with Virtualization Extensions \cite{kvm:2014}.
KVM/ARM introduces split-mode virtualization design,
in which the hypervisor are split into \textit{lowvisor} and \textit{highvisor}.
Lowvisor runs on Hyp mode to set up system registers, trap guest OS instructions, and handle interrupts.
Highvisor runs on non-Secure PL1 mode to leverage high-level functions of a Linux host.

Although these hypervisors are available in ARM architectures, they depend on Linux.
Some bugs on Linux may influence these hypervisors.
On the other hand, T-Visor does not depend on GPOSes.

In addition, they do not have real time schedulers.
However, there are existing approach to enable real-time scheduling in these hypervisors.
Yu et al. \cite{yu:2010} have improved real-time performance of the Xen scheduler.
Their new scheduler aware real-time guest OSes and support priority preemption.
PARFAIT \cite{parfait:2011} is a real-time scheduling architecture for Xen.
This scheduling architecture focus on guaranteeing CPU bandwidth for RTOSes.
RT-Xen \cite{rt-xen:2011, rt-xen-multi:2014} extends Xen to provide a real-time scheduling framework.
This framework enables to implement various scheduling algorithms in Xen.
Scheduling of KVM depends on a scheduler in Linux.
Therefore, the patches for Linux like PREEMPT-RT \cite{rt-linux} may improve real-time performance of KVM.

Quest-V \cite{quest-v:2014} is a hypervisor for embedded systems.
It partitions all physical resources at boot time to avoid traps into the hypervisor.
Their implementation does not support device sharing.
This approach is similar to our device manage policy.
However, it adopts its original scheduling algorithm and does not consider about selectability.
Moreover, its implementation is for x86-based architectures.

Before ARM Virtualization Extensions became available, paravirtualization approach are often proposed.
Suzuki et al. \cite{suzuki:2013} analyzed ARM architecture and constructed paravirtualization hypervisor on ARM architecture.
Their evaluation showed their hypervisor suffered from various overheads.
Paravirtualization approach needs to modify kernels.
This increases engineering costs.
If developers want to use new versions of kernels or other types of OSes,
they need to re-implement patches for them.

Except paravirtualization, various solutions were proposed for co-existence of multiple OSes in ARM architecture.
SafeG \cite{safeg:2010} is one of them.
This uses ARM Security Extension to run two OSes on the same board.
Modified Linux is running in the non-secure world, and an RTOS is running in the secure world.
Security Extensions is more popular than Virtualization Extensions.
Therefore, this approach can be applied to more widely platforms than ours.
Jing et al. \cite{trustzone:2013} also proposed a similar architecture using Security Extensions.
However, these methods require to modify OSes.
Moreover, these cannot run more than two OSes concurrently.

Aashish et al.\cite{aashish:2013} proposed a Type-1 hypervisor using DBT approach on the Power architecture.
The binary translation approach does not require any hardware extensions.
Therefore, this approach will be widely applied to other embedded architectures.
However, the engineering cost and execution overheads become bigger than trap and emulate hypervisor.

Microkernel design hypervisors are often proposed for embedded systems.
They leverage similarity between microkernels and hypervisors.
For example, OKL4 Microvisor \cite{okl4:2010} is a variant of the L4 microkernel.
This works on ARM processors and has been deployed to a lot of devices, mainly mobile phones.
It is called microvisor because it has features of both microkernels and hypervisors.
Microkernel design may be more secure than our design
because it separates the functions of the kernel into small processes.
NOVA \cite{nova:2010}, which is not for embedded systems,
focuses on how to implement a secure hypervisor and adopts a microkernel-based architecture.
There is an existing approach to apply formal verification to L4-based kernel \cite{sel4:2009}.
Such an approach will be applied to microkernel-based hypervisors more easily than our design.
However, it is also known that microkernels involve a lot of interprocess communications which make systems slower.
It is worthwhile to consider other types of hypervisor designs.

\subsection{Lightweight VMs}
A VM of hypervisors includes a whole OS.
It consumes as much memory and disk space as a normal OS.
Emulation of a whole hardware incurs execution overheads.
To moderate these costs, there are some approaches to provide VMs without hardware level virtualization.

Container-based virtualization, also called operating-system-level virtualization, is considered as a promising approach as a substitute for hardware level virtualization by hypervisor,
for example Linux-VServer \cite{vserver:2007}, LXC \cite{lxc}, FreeBSD jail \cite{jail:2000}.
This technique provides a shared OS images and system libraries.
To achieve security isolation, a host OS provides virtualized internal OS objects.
Each VM uses its own internal OS components.
Unlike hypervisor, container-based virtualization does not require any special hardware extension.
A size of the VM is smaller because OS kernel is shared.
Execution overheads are lower because it does not need to emulate sensitive instructions.
Cells \cite{cells:2011} is an example of container-based virtualization applied to mobile system.

Drawbridge \cite{drawbridge:2011} is an approach to construct light VMs without virtualization hardwares,
which is different from container-based virtualization.
This approach provides picoprocessess to each VM and Windows library OSes run on these picoprocesses.
Drawbridge aims to provide secure independent environments.
This approach does not need to emulate whole hardwares or to have whole OS images.
Therefore, it achieves low execution overheads and low memory usage.

However, container-based virtualization and Drawbridge require host OSes.
Container-based virtualization is mainly implemented in general-purpose OSes.
Drawbridge is implemented in Windows.
Therefore, these rely on a large TCB.
In addition, as a guest and host OS must be the same,
the developers cannot combine heterogeneous OSes.
Note, these approaches can combine with hardware virtualization.
For example, we can use Linux with LXC on T-Visor.

\subsection{Provide VMs without Traditional Hypervisors}
Exokernel \cite{exokernel:1995} is an OS architecture proposed by Engler et al.
In Exokernel, a kernel provides secure multiplexing of physical resources and applications manage their own resources.
This design enables applications to use physical resources by their optimized way and reduce overheads to access physical resources.
Each application runs on a library OS \cite{libos:1992}.
This OS runs in the same address space as an application runs.
Exokernel does not trust library OSes.
Therefore, faults of applications do not affect other applications.
The scheduling in Exokernel is also flexible.
However, this architecture cannot run traditional OSes directly.

NoHype \cite{nohype:2010} is an approach to construct virtualized cloud infrastructure
without the conventional virtualization layer.
In this approach, each processor core is dedicated to a single VM.
Therefore, we cannot run more VMs than the processor cores.
Although multi-core processors are common in embedded systems, it seems inconvenient.
In addition, it assumes that all devices are virtualized in hardware level.
This assumption strongly restricts platforms.

\section{Future Work}
\subsection{Toward Multi-core Support}
We are working on supporting multi-core environments.
We explain some of main features to implement.

First, we need to extend the virtual GIC Distributor.
Some of the resisters in the distributor have different values among cores.
We also need to implement support for software generated interrupts.
In uniprocessor systems, these functions are not used.
There are peripheral interrupts that target more than one processor.
In bare-metal environments, after one of targeted processors receives such an interrupt by reading the acknowledge register, other cores receive spurious interrupts so that only one core receives the interrupt.
However, virtual CPU interfaces do not have such mechanisms to prevent virtual peripheral interrupts from being received by more than one processor.
Therefore, virtual Distributor must decide which virtual core should receive peripheral interrupts.
We have implemented some of these features, but they are not perfect.

\texttt{wfe} instruction may be a problem.
This instruction lets a processor sleep until \texttt{sev} is called in other processors.
We can capture \texttt{wfe}.
However, there is no way to capture \texttt{sev} instructions in ARMv7-A architecture.
This instruction is often used to implement spin lock in multi-core environments.
Most programs are implemented to work correctly if \texttt{wfe} does not block the execution.
Therefore, T-Visor emulates \texttt{wfe} as \texttt{nop} in current implementation.

In multi-core system software for ARM architecture,
they usually use Power State Coordination Interface (PSCI).
Through this common interface, system softwares let other CPU start or let their own CPU suspend.
Its implementation is usually in the boot loader.
To access PSCI, system software must use a Hyp call or a Secure Monitor call (SMC).
Hypervisors can capture both Hyp call and SMC.
We are now implementing PSCI to control virtual CPUs of T-Visor.

A case study of our flexible scheduling framework was shown in this paper.
However, we need to evaluate our scheduling framework for multi-core environments.
In multi-core environments, scheduling algorithms are more diverse and complex.
For example, how to assign virtual cores to physical cores is one of difficult problems.

\subsection{Smart Configuration}
We showed our scheduling framework.
In this framework, they need to implement their desired algorithms in C or assembly without any standard libraries.
We provide some data structures of our original library.
However, it is not enough to support implementation.
More high-level language, such as C++, may be also helpful to implement not only schedulers but also T-Visor itself.

In addition, developers must configure how to allocate physical resources to VMs by writing C code.
This may lead to some troubles.
We are considered to design Domain Specific Language (DSL) or configuration scripts.

\subsection{Compatibility}
Recently, unikernels \cite{unikernel:2013} become attractive in the cloud environments.
This type of OSes are optimized for specific application like library OS.
They do not have several abstraction support which traditional OSes have, such as process.
Unikernels are designed to run on hypervisors, so that they employ virtual devices.
This makes unikernels portable between different hardwares.
Contrast to traditional OSes, their binary size and execution overheads in applications are low.

We consider that unikernels are useful in the embedded systems too.
However, these unikernels assume general hypervisors.
They use virtual devices.
To run unikernels, we need to implement compatible virtual devices in T-Kernel.
However, it is hard to implement virtual devices from scratch.
We are considering to use Rump Kernel \cite{rumpkernel:2015} to provide virtual devices.
This provides device drivers of NetBSD.
But we do not decide how to integrate Rump Kernel into T-Visor.

\subsection{Security}
TCB of T-Visor is much smaller than other hypervisors because T-Visor does not rely on GPOSes or large libraries.
However, there are still possibility that T-Visor includes bugs or security vulnerabilities.
One way to confirm the safety of our system is formal verification.
Dam et al. attempted to verify an ARM hypervisor \cite{verification_arm:2013}.
The small code of T-Visor may enable these approaches.

Another way is testing.
Amit et al. \cite{vcpu:2015} have proposed the validation of hypervisors using the testing environment of CPU vendors.
Their approach is testing VCPUs of KVM on Intel processor using Intel's testing facility for physical CPUs.
This approach will be useful for T-Visor to improve its reliability.

\section{Conclusions}
We showed the design and implementation of T-Visor,
a hypervisor oriented for mixed criticality embedded real-time system.
We employed hardware virtualization support to realize simple and fast virtualization.
To satisfy various real-time requirements, we designed a flexible scheduling framework of T-Visor.
Our implementation is much smaller than general-purpose hypervisors.
Our experiments showed T-Visor could run GPOS with lower overheads than a general purpose hypervisor.

As we mentioned previously, we are working on supporting multi-core environments.
In multi-core environments, scheduling algorithms are more diverse and complex.
In future work, we plan to re-consider our scheduling framework of multi-core environments.
In addition, we plan to improve other implementation, such as device handling.
And then, we will apply T-Visor to real world applications.

T-Visor is an open source product.
Our implementation is available in \url{https://github.com/garasubo/T-Visor}.


%
\bibliographystyle{abbrv}
\bibliography{refs}  
%
%
\end{document}